\documentclass[runningheads]{llncs}
\usepackage[T1]{fontenc}
\usepackage{amsmath}
\usepackage{float}
\usepackage{orcidlink}
\usepackage{physics}
\usepackage{subcaption}
\begin{document}
\title{Learning Robust Observable to Address\\Noise in Quantum Machine Learning}
\titlerunning{Learning Robust Observable to Address Noise in QML}
\author{Bikram Khanal~\orcidlink{0000-0003-2292-520X} \and
Pablo Rivas~\orcidlink{0000-0002-8690-0987}}
\authorrunning{B. Khanal et al.}

\institute{Department of Computer Science, Baylor University, Texas, USA
\email{\{Bikram\_Khanal1,Pablo\_Rivas\}@Baylor.edu}}
\maketitle              
\begin{abstract}
Quantum Machine Learning (QML) has emerged as a promising field that combines the power of quantum computing with the principles of machine learning. One of the significant challenges in QML is dealing with noise in quantum systems, especially in the Noisy Intermediate-Scale Quantum (NISQ) era. Noise in quantum systems can introduce errors in quantum computations and degrade the performance of quantum algorithms. In this paper, we propose a framework for learning observables that are robust against noisy channels in quantum systems. We demonstrate that it is possible to learn observables that remain invariant under the effects of noise and show that this can be achieved through a machine-learning approach. We present a toy example using a Bell state under a depolarization channel to illustrate the concept of robust observables. We then describe a machine-learning framework for learning such observables across six two-qubit quantum circuits and five noisy channels. Our results show that it is possible to learn observables that are more robust to noise than conventional observables. We discuss the implications of this finding for quantum machine learning, including potential applications in enhancing the stability of QML models in noisy environments. By developing techniques for learning robust observables, we can improve the performance and reliability of quantum machine learning models in the presence of noise, contributing to the advancement of practical QML applications in the NISQ era.

\end{abstract}
\section{Introduction}
The intersection of quantum computing and machine learning has given rise to the rapidly evolving field of QML, offering new paradigms for solving complex computational problems~\cite{trugenberger2002quantum,wittek2014quantum}. QML aims to leverage quantum mechanics properties such as superposition and entanglement to enhance the capabilities of traditional machine learning algorithms~\cite{schuld2019evaluating}. Various quantum machine learning algorithms~\cite{lloyd2010quantum,cerezo2021variational,lloyd2014quantum,maheshwari2021variational,abbas2021power} and optimization techniques~\cite{biamonte2017quantum,Rebentrost2018Quantum,Bittel2021Training,schuld2019evaluating} have been proposed for solving machine learning tasks~\cite{biamonte2017quantum,havlivcek2019supervised,liu2021rigorous,rivas2024quantum}. However, one of the major challenges in QML is dealing with the presence of noise in quantum systems, especially in the Noisy Intermediate-Scale Quantum (NISQ) era~\cite{khanal2023evaluating,preskill2018quantum,khanal2024modified}. 

Noise in quantum systems can arise from various sources, including imperfect gates, decoherence, and environmental interactions, which introduce errors in quantum computations and degrade algorithm performance~\cite{nielsen2002quantum,preskill2018quantum}. In the context of QML, such noise can affect the training and inference processes of quantum machine learning models, leading to suboptimal results~\cite{schuld2019quantum}. These challenges extend beyond algorithm design, impacting the fundamental principles of QML. Specifically, the inherent noise in NISQ machines poses significant obstacles to the learning capabilities of Quantum Neural Networks~\cite{Du2021learnability}. Additionally, system noise can significantly reduce the quantum kernel advantage, raising concerns about the feasibility of quantum kernel methods~\cite{huang2021power,wang2021towards}. Furthermore, computing numerical gradients on noisy qubits requires a delicate balance; reducing the step size to enhance accuracy can obscure subtle differences in the cost function for nearby parameter values~\cite{pop00021}.

Several techniques have shown theoretical promise in enhancing the accuracy and robustness of QML models in the presence of noise. These techniques include error mitigation~\cite{kandala2019error}, quantum error correction~\cite{shor1995scheme}, variational quantum thermalizing algorithm~\cite{mcardle2019variational}, zero noise extrapolation~\cite{temme2017error}, and randomized circuit resampling~\cite{temme2017error}. Studying quantum variational classification in the presence of noise is not only of theoretical significance but also of practical importance. As NISQ devices evolve, understanding the noise effects on quantum models becomes crucial for efficient quantum algorithm and hardware design~\cite{havlivcek2019supervised}. One particular area of interest is the robustness of observables in quantum systems against noise. Observables are essential components of quantum measurements and are crucial in training machine learning models~\cite{nielsen2002quantum,schuld2021machine}. Understanding the robustness of observables against noise can provide valuable insights into the stability and reliability of the QML model.

In this paper, we propose a framework for learning observables that are robust against noisy channels in quantum systems. We demonstrate that it is possible to learn robust observables that remain invariant under the effects of noise, and we show how this can be achieved through a machine-learning approach. Furthermore, we provide a toy example to illustrate the concept of robust observables and then describe a machine-learning framework for learning such observables. Our result shows that we can train the quantum circuits under noise to learn observables that are more robust to noise than conventional observables. We discuss the implications of this finding for quantum machine learning, including potential applications in enhancing the stability of QML models in noisy environments. By developing techniques for learning robust observables, we can enhance the performance and reliability of quantum machine learning models in the presence of noise, contributing to the advancement of practical QML applications in the NISQ era.
The main contributions of this paper are as follows:
\begin{enumerate}
    \item Despite the varied noise rates and channels, there are observables for each state-channel combination where the expectation value remained constant. 
    \item This result is unexpected because conventional literature suggests that noise should generally degrade quantum information, leading to varying expectation values.
    \item Observables effectively ``filters out'' the noise under the measurement process,
     which exhibits the robustness property of the observables, at least within the range of noise rate $p$.
\end{enumerate}

\section{Background}
The NISQ-era quantum system is subject to various noise sources that can introduce errors and degrade the performance of quantum algorithms. These noise sources can be modeled as quantum channels that act on the quantum state and introduce errors in the system. This section will define some common noise channels that model noise in quantum systems.

\subsubsection{Depolarizing Channel}
The depolarizing channel is a noise channel that occurs in quantum systems due to the loss of coherence and the introduction of errors. The depolarizing channel is represented by a completely positive trace-preserving (CPTP) map that acts on the density matrix $\rho$ of the quantum state as:
\begin{equation} \label{eq:depolarizing_channel}
    \rho \rightarrow (1-p)\rho +  p \frac{I}{d} ,
\end{equation}
where $p$ is the depolarization rate, $I$ is the identity matrix, and $d$ is the dimension of the quantum system. The depolarizing channel introduces errors in the quantum state by replacing the state with a completely mixed state with probability $p$.

\subsubsection{Amplitude Damping Channel}
The amplitude damping channel is another noise channel that occurs in quantum systems due to energy loss and quantum state decay. The amplitude damping channel is represented by a CPTP map that acts on the density matrix $\rho$ of the quantum state as:
\begin{equation} \label{eq:amplitude_damping_channel}
    \rho \rightarrow E_0 \rho E_0^\dag + E_1 \rho E_1^\dag ,
\end{equation}
where $E_0$ and $E_1$ are given as:
\begin{equation}
    E_0 = \begin{pmatrix}
    1 & 0 \\
    0 & \sqrt{1-\gamma}
    \end{pmatrix}, \quad
    E_1 = \begin{pmatrix}
    0 & \sqrt{\gamma} \\
    0 & 0
    \end{pmatrix} ,
\end{equation}
with $\gamma$ being the damping rate. 

\subsubsection{Phase Damping Channel}
The phase damping channel is a noise channel that occurs in quantum systems due to the loss of phase information and the introduction of errors. The phase damping channel is represented by a CPTP map that acts on the density matrix $\rho$ of the quantum state as:
\begin{equation} \label{eq:phase_damping_channel}
    \rho \rightarrow E_0 \rho E_0^\dag + E_1 \rho E_1^\dag ,
\end{equation}
where $E_0$ and $E_1$ are given as:
\begin{equation}
    E_0 = \begin{pmatrix}
    1 & 0 \\
    0 & \sqrt{1-\gamma}
    \end{pmatrix}, \quad
    E_1 = \begin{pmatrix}
    0 & 0 \\
    0 & \sqrt{\gamma}
    \end{pmatrix} ,
\end{equation}
with $\gamma$ being the phase damping rate.

\subsubsection{Phase Flip Channel}
The phase flip channel is a noise channel that occurs in quantum systems due to the introduction of phase errors. The phase flip channel is represented by a CPTP map that acts on the density matrix $\rho$ of the quantum state as:
\begin{equation} \label{eq:phase_flip_channel}
    \rho \rightarrow (1-p) \rho + p Z \rho Z ,
\end{equation}
where $p$ is the phase flip rate and $Z$ is the Pauli-Z matrix. The phase flip channel introduces phase errors in the quantum state with probability $p$.

\subsubsection{Bit Flip Channel}
The bit flip channel is a noise channel that occurs in quantum systems due to the introduction of bit errors. The bit flip channel is represented by a CPTP map that acts on the density matrix $\rho$ of the quantum state as:
\begin{equation} \label{eq:bit_flip_channel}
    \rho \rightarrow (1-p) \rho + p X \rho X ,
\end{equation}
where $p$ is the bit flip rate and $X$ is the Pauli-X matrix. The bit flip channel introduces bit errors in the quantum state with probability $p$.

\subsubsection{Quantum Observables}
Observables are an essential concept in quantum mechanics that describe the physical quantities that can be measured in a quantum system. Observables are represented by Hermitian operators that act on the quantum state and correspond to the physical properties that can be observed in the system. In quantum mechanics, observables describe the outcomes of measurements and the probabilities of different measurement results. Observables play a crucial role in quantum algorithms and machine learning models, where they are used to extract information from quantum states and perform quantum measurements.

\section{Problem Definition}
To overcome the challenge of noise in quantum systems, we aim to develop a framework for learning observables that are robust against noisy channels. We focus on learning observables that remain invariant under the influence of noise in quantum systems. Specifically, `observables that remain invariant' refers to quantum observables whose expectation values do not change despite the presence of noise channels. This invariant is a crucial property for maintaining the stability and reliability of measurements in quantum systems. In mathematical terms, an observable $\mathcal{O}$ is considered invariant under a noise model ${E}$ represented by Kraus operators $K_i$ if and only if each $K_i$  commutes with $O$, i.e. $K_i^\dag O K_i = O$ for all $i$ (details in~\ref{thm:invariant_expectation}). Practically, this means that the noise the quantum channel introduces does not affect the measurement outcomes associated with these observables.

We take a machine-learning approach to learn these observables by training QML models on noisy quantum systems and optimizing the observables to be robust against noise. By learning robust observables, we can enhance the stability and accuracy of quantum machine learning models and improve their performance in the presence of noise. Next, we define a theorem that provides a necessary and sufficient condition for the invariance of an observable expectation value under a noisy channel.

\begin{theorem} \label{thm:invariant_expectation}
    The expectation value $\expval{O}$ of an observable $O$ on a quantum state $\rho$ remains invariant under a noise model $\mathcal{E}$, represented by Kraus operators $\{K_i\}$, if and only if each $K_i$ commutes with $O$, i.e., $K_i^\dag O K_i = O$ for all $i$.
    \begin{proof}
    Let $O$ be an observable and $\rho$ be a $d$-dimensional quantum state. For a general quantum noise channel $\mathcal{E}$ represented by Kraus operators $\{K_i\}$, the expectation value of the observable $O$ under the noisy channel is given by:
\begin{equation}
    \expval{O}_{\mathcal{E}(\rho)} = \text{Tr}(O \mathcal{E}(\rho)) = \text{Tr}\left(O \sum_{i} K_i \rho K_i^\dag \right) .
\end{equation}
Using the cyclic property of the trace, we can write the expectation value as:
\begin{equation}
    \expval{O}_{\mathcal{E}(\rho)} = \sum_{i} \text{Tr}(K_i^\dag O K_i \rho) .
\end{equation}
From the assumption of the invariant expectation value, we require that:
\begin{equation}
    \sum_i \text{Tr}(K_i^\dag O K_i \rho) = \text{Tr}(O \rho) .
\end{equation}
The above equation must hold for all states $\rho$, which implies that:
\begin{equation}\label{eq:invariant_expectation}
    \sum_i K_i^\dag O K_i = O .
\end{equation}
Next, we need to show that $\sum_i K_i^\dag O K_i = O$ is equivalent to $K_i^\dag O K_i = O$ for all $i$.

\subsection*{Sufficiency}
First, assume $K_i^\dag O K_i = O$ for all $i$. Then we have:
\begin{equation}
    \sum_i K_i^\dag O K_i = \sum_i O = O .
\end{equation}
Thus, $\sum_i K_i^\dag O K_i = O$ is satisfied.

\subsection*{Necessity}
Now assume $\sum_i K_i^\dag O K_i = O$. Consider the linearity of the trace:
\begin{equation}
    \text{Tr}\left(\left(\sum_i K_i^\dag O K_i\right) \rho\right) = \text{Tr}(O \rho) ,
\end{equation}
for all $\rho$, which implies:
\begin{equation}
    \sum_i K_i^\dag O K_i = O .
\end{equation}
Since this must hold for any observable $O$, each term in the sum must individually satisfy $K_i^\dag O K_i = O$.

Therefore, the condition $\sum_i K_i^\dag O K_i = O$ is equivalent to the condition that each $K_i$ commutes with $O$, i.e., $K_i^\dag O K_i = O$ for all $i$.
\end{proof}
\end{theorem}
The above theorem provides a necessary and sufficient condition for the invariance of an observable expectation value under a noisy channel. In the next section, we will discuss how we can leverage the machine learning approach to learn observables that satisfy this condition and are robust against noisy channels.
\section{Result}

\subsection{Problem Example}
This section provides a toy example to investigate the impact of the noisy channel on quantum states' expectation values. In particular, we want to understand if any observables are more robust to noise than others. We consider an observable $O$ more robust than observable $O'$ if the expectation value of $O$ diverts less from the ideal expectation value as the noise rate increases.

For example, consider the Bell state, $B_{++}$. The state vector for this Bell state can be written as:
\begin{equation}\label{eq:bell_state}
    \ket{\Phi^+} = \frac{1}{\sqrt{2}}(\ket{00} + \ket{11}) .
\end{equation}
Which can be represented in a density matrix form as:
\begin{equation}\label{eq:rhobell}
    \begin{split}
    \rho & = \ket{\Phi^+}\bra{\Phi^+} \\
    & = \frac{1}{2}\begin{pmatrix}
    1 & 0 & 0 & 1 \\
    0 & 0 & 0 & 0 \\
    0 & 0 & 0 & 0 \\
    1 & 0 & 0 & 1
    \end{pmatrix} .
    \end{split} 
\end{equation}

We can compute the expectation value of an observable $O$ on the Bell state using the formula:
\begin{equation}\label{eq:expectation_value_ideal}
    \expval{O} = \text{Tr}(O\rho) .
\end{equation}
Eq.~(\ref{eq:expectation_value_ideal}) gives the expectation value of the observable $O$ in an ideal situation. 
Next, we consider the depolarizing channel~\ref{eq:depolarizing_channel} to simulate the noise. We will then measure the expectation value of an observable for the depolarized Bell state and compare it with the ideal expectation value.

As a part of the toy example, let us simulate (\ref{eq:rhobell}) under the depolarization channel~\ref{eq:depolarizing_channel} and compute the density matrix for the depolarized Bell state as:

\begin{equation}\label{eq:rho_depolarized}
    \begin{split}
    \rho_{depolarized} & = (1-p)\rho +  p \frac{I}{4} \\
    & = \frac{1}{2}\begin{pmatrix}
    1 - \frac{p}{2} & 0 & 0 & 1- p \\
    0 & \frac{p}{2} & 0 & 0 \\
    0 & 0 & \frac{p}{2} & 0 \\
    1-p & 0 & 0 & 1 - \frac{p}{2}
    \end{pmatrix} .
    \end{split}
\end{equation}
We then make a measurement using any observables on~(\ref{eq:rho_depolarized}) to get the expectation value of the observable. It is widely known that as we increase the depolarization rate, the expectation value of the observables will change. We can use this example to investigate if there are observables that are more robust to noise than others. 

Let us construct an arbitrary observable $O_{optimized}$ given by~(\ref{eq:toy_obs}).
\begin{figure*}[h]
\begin{equation}\label{eq:toy_obs}
    O_{optimized} = \begin{pmatrix}
    0.804 & 0.086 + 0.138i & 0.739 + 0.050i & 0.070 + 0.132i \\
    0.086 - 0.138i & 0.302 & 0.087 - 0.122i & 0.277 + 0.019i \\
    0.739 - 0.050i & 0.087 + 0.122i & 1.253 & 0.133 + 0.215i \\
    0.070 - 0.132i & 0.277 - 0.019i & 0.133 - 0.215i & 0.470
    \end{pmatrix} .
\end{equation}
\end{figure*}

We can compute its expectation value on~(\ref{eq:rho_depolarized}) using:
\begin{equation}\label{eq:expectation_value}
    \expval{O_{optimized}} = \text{Tr}(O_{optimized}\rho_{depolarized}) .
\end{equation}

\begin{figure}[h!]
    \centering
    \includegraphics[width=0.8\textwidth]{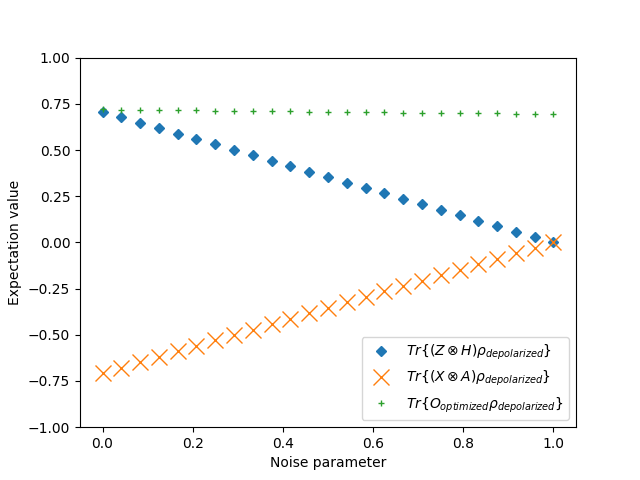}
    \caption{Expectation value of different observables on the depolarized Bell state as a function of the depolarization rate $p$. $Z$ is the Pauli-Z matrix, $X$ is the Pauli-X matrix, $H$ is the Hadamard gate, and $O_{optimized}$ is an arbitrary single qubit Hermitian measurement operator. The expectation value of the observable $O_{optimized}$ remains constant as the depolarization rate $p$ increases.}
    \label{fig:expectation_value_bell_plus}
\end{figure}

We then compare the expectation values of $O_{optimized}$ on the depolarized Bell state to those on the ideal Bell state, considering the depolarization rate $p \in [0,1)$for this example. The expectation values of the observable $O_{optimized}$ on the depolarized Bell state as a function of the depolarization rate $p$ are plotted in Fig.~\ref{fig:expectation_value_bell_plus}. Our observations indicate that $\expval{O}$ for (\ref{eq:rho_depolarized}) remains consistent at approximately $0.70$ as the depolarization rate increases.

Fig.~\ref{fig:expectation_value_bell_plus} shows the intriguing result that the expectation value of the custom observable $O_{optimized}$ on the depolarized Bell state remains constant as the depolarization rate $p$ increases.
However, conventional observables, such as the Pauli matrices and Hadamard gate, are not robust to noise.
We use this example to motivate our investigation of whether some observables are more robust to noise channels than others.

\subsection{Learning Robust Observables}

Motivated by the example above, we want to investigate if learning observables robust to noise in an arbitrary circuit would be possible.
We considered six different circuits: four Bell states circuits, a two-qubit Quantum Fourier Transform (FFT) circuit, and a two-qubit highly entangled random circuit. 
Similarly, we considered the depolarization, amplitude damping, phase damping, phase flip, and bit flip noise channels.
We selected twenty-five different values between $0$ and $1$ at uniform intervals for noise rates. For each state-channel combination, the goal is to learn an observable that is robust to noise across all degrees. We briefly describe our approach below.

To begin with, we considered the Pauli-Z matrix to be an observable in an ideal situation. The computed expectation value of Pauli-Z in the ideal setting is then used as the target value for the learning process. For Each qubit, we randomly initialized a $2 \times 2 $ observables. We used these randomly generated observables to measure the expectation value of the circuit under the noisy channel. The circuit's expectation value under the noisy channel is then compared to the target value to calculate the difference between the ideal and noisy settings. In a machine learning framework, the expectation value under the ideal setting is a true label, i.e., $\expval{O}_{\text{ideal}} = y$, and the expectation value under each noisy channel is $\expval{O_i}_{\text{noisy}} = \hat{y}$. With these analogies, we used the standard absolute square loss function to compute the loss for each iteration. The cost function based on the absolute square loss function is given by:
\begin{equation}\label{eq:loss}
    C(\theta) =  \frac{1}{N}\sum_{i=0}^{N}\abs{\expval{O_i}_{\text{noisy}} - \expval{O}_{\text{ideal}}}^2 .
\end{equation}
Where $N$ is the total number of values in noise rate, twenty-five in this case, $\langle O_i \rangle$ is the expectation value
of the observable $O_i$ under the noisy channel, with $i$ being the index of the noise rate, and $\expval{O}$ is the expectation value of the observable $O$ in the ideal condition.
We used the parameter-shift rule to compute the cost function gradient with respect to the observable parameters. Each model for each state-channel combination is trained for $300$ epochs with a learning rate of $0.1$.
We can define the parameter-shift rule as:
\begin{equation}\label{eq:parameter_shift_rule}
    \nabla_{\theta}C = \frac{1}{2}\left(C(\theta + \frac{\pi}{2}) - C(\theta - \frac{\pi}{2})\right) .
\end{equation}
Where $\theta$ is the parameter of the observable, and $C(\theta)$ is the cost function with respect to the parameter $\theta$.

\begin{figure*}[h!]
    \centering
    \begin{subfigure}[b]{\textwidth}
        \centering
        \includegraphics[width=\textwidth]{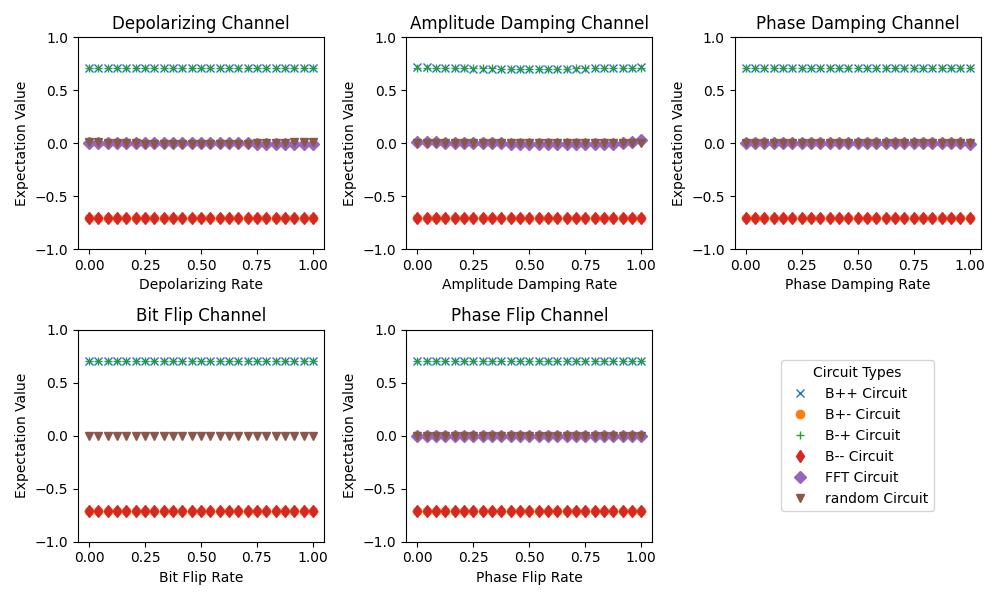}
        \caption{Expectation value of various observables on each circuit-channel combination. The x-axis is the noise rate for the corresponding channel, 
        and the y-axis is the expectation value of the observable. The different colors represent different circuits.}
        \label{fig:expectation_values}
    \end{subfigure}
    \hfill
    \begin{subfigure}[b]{\textwidth}
        \centering
        \includegraphics[width=\textwidth]{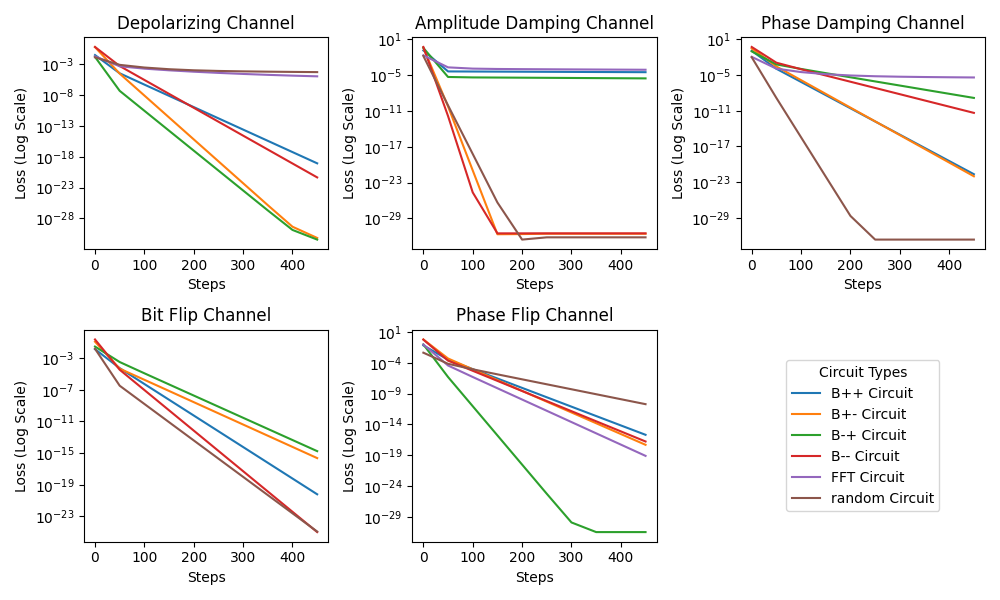}
        \caption{Training loss for each circuit-channel combination. The x-axis is the number of epochs, and the y-axis is the log scaling of the training loss.
        The different colors represent different circuits.}

        \label{fig:training_loss}
    \end{subfigure}
    \caption{Results of the learning process}
    \label{fig:learning_results}
\end{figure*}

\begin{figure*}[h!]
    \centering
    \includegraphics[width=\textwidth]{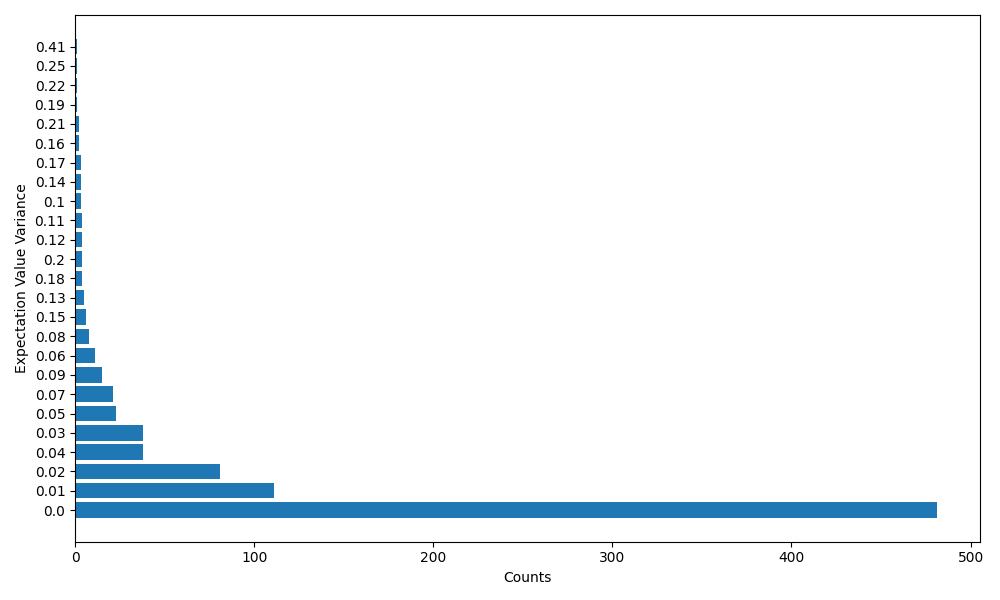}
    \caption{Count of the standard deviation of the expectation value of the observables for all the circuit-channel combinations.}
    \label{fig:standard_deviation}
\end{figure*}

Fig.~\ref{fig:learning_results} provides the performance matrices of various circuits 
under different noisy channels. The logarithmic scale on the y-axis of Fig.~\ref{fig:training_loss} 
indicates the exponential decay of the training loss as the number of epochs increases. The expectation values
of the observables for each circuit-channel combination are shown in Fig.~\ref{fig:expectation_values}.
The expectation value of the observables for each circuit-channel combination remains steady regardless of the noise rate. This might indicate the inherent symmetries or error-resistance properties within these circuits.
We trained the circuits to learn thirty observables for the thirty possible combinations of the circuits and channels. In the following sections, we will discuss some of the statistics and properties of the learned observables.

\subsection{Statistics of the Learned Observables}
We begin this section by analyzing the properties of the learned observables. For each circuit-channel combination, we learned two $2 \times 2$ observables corresponding to each qubit, named $O_1$ and $O_2$. We can then get the observables for the current system as $\mathcal{O} = \ O_1 \otimes O_2$. Some of the properties of the learned observables are:
\begin{itemize}
    \item All the observables are Hermitian. This is expected because the expectation value of an observable is real.
    \item All the observables have real eigenvalues.
    \item The eigenvectors of each observable are orthogonal to each other.
    \item Each observable's eigenvalues sum is equal to its trace.
    \item  Each observable is a linear combination of the Pauli matrices.
\end{itemize}

To analyze if the robust observable learned for one circuit-channel study is robust on a different circuit-channel combination, we computed the expectation value of each observable on each circuit-channel combination. To clarify, each observable was used to calculate the expectation value for all 30 independent circuit-channel combinations. 
In each case, we computed the expectation value of the observable for twenty-five different values of the noise rate. We then calculated the standard deviation of the expectation value for each condition. The standard deviation range across all the circuit channel combinations is $(0.0, 0.4068)$. To understand it better, we plot the bar chart to count the standard deviation of expectation values. Fig.~\ref{fig:standard_deviation} shows that most observables, almost $500$, have a standard deviation of zero. This indicates that the expectation value of the observables for the majority of the circuit-channel combinations remains constant to the noise rate, implying that the observables are robust to noise across most of the circuit-channel combinations. This result is consistent with the toy example we discussed earlier. Hence, we argue that learning observables robust to noise in quantum systems is possible. The learned observables can provide reliable measurements in noisy quantum systems and enhance the stability and accuracy of quantum machine learning models.

\section{Discussion}
Developing robust QML techniques remains a critical challenge as we progress through the NISQ era. While classical machine learning has demonstrated remarkable capabilities across diverse domains~\cite{goodfellow2016deep,rajkomar2019machine,senior2020improved}, translating these successes to quantum systems introduces unique obstacles stemming from inherent noise and decoherence effects~\cite{preskill2018quantum,khanal2023evaluating,arute2019quantum,krantz2019quantum}. The pursuit of noise-resilient QML approaches is thus of paramount importance for realizing practical quantum advantages in the near term.

Identifying and utilizing quantum observables that exhibit robustness against common noise channels is a promising approach. By leveraging noise-invariant observables, QML models may maintain their performance even in environmental perturbations. Unlike traditional error mitigation strategies, it does not require additional quantum resources or circuit depth, making it well-suited for near-term devices with limited coherence times. Also, addressing noise at the measurement level complements existing error correction protocols that operate during computation. The results of our toy example and machine learning framework demonstrate the feasibility of learning noise-robust observables in quantum systems. 

However, the identification of noise-robust observables presents its own set of challenges. The space of possible observables grows exponentially with system size, making exhaustive search intractable for all but the smallest quantum systems. Furthermore, the relationship between an observable structure and noise resilience is not immediately apparent.
Using machine learning techniques to discover noise-robust observables represents an intriguing synergy between classical and quantum computation. This hybrid approach leverages classical optimization algorithms to navigate the vast space of potential observables, guided by quantum measurements of their performance under various noise models. Such a methodology aligns well with the current limitations of NISQ devices, allowing for iterative improvement of quantum algorithms through classical post-processing. Generalizing learned observables across different quantum states and noise models requires further investigation. It is conceivable that observables optimized for specific noise channels may not maintain their robustness under different error processes, especially when we increase the number of qubits. Future work will explore the generalization capabilities of learned observables and their applicability to a broader range of quantum systems.

\section{Conclusion}
The pursuit of noise-robust observables is a promising field for enhancing the reliability of QML in the NISQ era. In this paper, we proved that learning observables that remain invariant under the effects of noise in quantum systems is possible. We demonstrated that the expectation value of the observables remains invariant under the impact of noise and that we may leverage a machine-learning approach to learn such observables. The key idea is to learn the observables that are insensitive to noise and can provide reliable measurements in noisy quantum systems. By addressing noise at the measurement level, our approach complements existing error mitigation techniques and aligns well with the current capabilities of quantum hardware. We presented a toy example to illustrate the concept of robust observables and then described a machine-learning framework for learning such observables. By developing techniques for learning robust observables, we can enhance the performance and reliability of quantum machine learning models in the presence of noise. Future work will extend this framework to more complex quantum systems and explore robust observables' applications in quantum machine learning models. As the field progresses, overcoming scalability challenges and deepening our theoretical understanding will be crucial for realizing the full potential of this methodology in practical QML applications.

\begin{credits}
\subsubsection{\ackname} Part of this work was funded by the National Science Foundation under grants CNS-2210091, CHE-1905043, and CNS-2136961. The authors also acknowledge the support of the Baylor AI lab at Baylor University's Department of Computer Science.

\subsubsection{\discintname}
The authors have no competing interests to declare that are relevant to the content of this article.
\end{credits}

\bibliographystyle{splncs04}
\bibliography{ref}
\end{document}